\documentclass[twocolumn,
showpacs,preprintnumbers,amsmath,amssymb]{revtex4}

\usepackage{amssymb,amsmath,amsthm}
\usepackage{epsfig}
\usepackage{multirow}

\theoremstyle{definition}

\newcommand{\bra}[1]{{\left\langle #1 \right|}}
\newcommand{\ket}[1]{{\left| #1 \right\rangle}}

\newcommand{\B}{\mbox{$\mathbb B$}}
\newcommand{\X}{\mbox{$\mathbb X$}}

\newcommand{\T}{\mbox{$\mathrm{tr}$}}

\begin{document}
\title{Strong polygamy of quantum correlations in multi-party quantum systems}

\author{Jeong San Kim}
\email{freddie1@suwon.ac.kr} \affiliation{
 Department of Mathematics, University of Suwon, Kyungki-do 445-743, Korea
}
\date{\today}

\begin{abstract}
We propose a new type of polygamy inequality for multi-party quantum
entanglement. We first consider the possible amount of bipartite entanglement distributed
between a fixed party and any subset of the rest parties in a multi-party quantum system.
By using the summation of these distributed entanglements, we provide an upper bound of the
distributed entanglement between a party and the rest in multi-party quantum systems.
We then show that this upper bound also plays as a lower bound of the usual polygamy inequality,
therefore the strong polygamy of multi-party quantum
entanglement. For the case of multi-party pure states, we further show
that the strong polygamy of entanglement implies the strong polygamy of quantum discord.
\end{abstract}

\maketitle
\section{Introduction}

Entanglement is one of the most remarkable features in the field of quantum
information and computation theory with many useful
applications such as quantum teleportation and quantum key distribution~\cite{tele,BB84,Eke91}.
One of the essential differences of quantum entanglement from other
classical correlations is in its restricted shareability; if a pair of parties in a multi-party quantum system
share maximal entanglement, they cannot have any entanglement nor
classical correlations with the rest. This restricted shareability
of entanglement is known as the {\em monogamy of entanglement}
(MoE)~\cite{T04, JGS}, which does not have any classical counterpart, and this makes
quantum physics fundamentally different form classical physics.

In the seminal paper by Coffman, Kundu and Wootters~\cite{CKW}, MoE
was mathematically characterized in forms of a trade-off inequality; for a
three-qubit state $\rho_{ABC}$ with two-qubit reduced density
matrices $\rho_{AB}=\T_C \rho_{ABC}$  and $\rho_{AC}=\T_B
\rho_{ABC}$,
\begin{align}
\tau\left(\rho_{A(BC)}\right)\geq \tau\left(\rho_{AB}\right)+\tau\left(\rho_{AC}\right),
\label{CKW}
\end{align}
where $\tau\left(\rho_{A(BC)}\right)$ is the entanglement of
$\rho_{ABC}$ with respect to the bipartition
between A and BC measured by tangle~\cite{CKW}, and $\tau\left(\rho_{AB}\right)$ and
$\tau\left(\rho_{AC}\right)$ are the tangles of $\rho_{AB}$ and
$\rho_{AC}$ respectively. Later, Inequality~(\ref{CKW}) was
generalized for multi-qubit systems~\cite{OV} and some classes
of multi-qudit systems in terms of various entanglement
measures~\cite{Kim123}. It was recently shown that squashed
entanglement is a faithful entanglement measure, which also shows
the monogamy inequality of entanglement in arbitrary dimensional
quantum systems~\cite{BCY10}.

Whereas MoE shows the restricted shareability of bipartite
entanglement in multi-party quantum systems, the possible amount of
bipartite entanglement distribution assisted by the third party is
known to have a dually monogamous (thus  polygamous) property in
multi-party quantum systems; for three-qubit systems, {\em
polygamy of entanglement} (PoE) was first characterized as a polygamy inequality
\begin{align}
\tau\left(\rho_{A(BC)}\right)\leq
\tau_a\left(\rho_{AB}\right)+\tau_a\left(\rho_{AC}\right),
\label{poly1}
\end{align}
where $\tau_a\left(\rho_{AB}\right)$ and
$\tau_a\left(\rho_{AC}\right)$ are the tangle of assistance of
$\rho_{AB}$ and $\rho_{AC}$ respectively~\cite{GMS, GBS}.
Inequality~(\ref{poly1}) was generalized for various classes of
multi-party, higher dimensional quantum systems~\cite{BGK}, and
a general polygamy inequality of entanglement was recently shown in
terms of {\em entanglement of assistance} in arbitrary
dimensional multi-party quantum systems~\cite{gpoly}.

The study of shareability and distribution of quantum correlations,
especially quantum entanglement, in multi-party quantum systems is
the key ingredient of many quantum information and communication
protocols. For example, due to the mutually-exclusive relation of
entanglement sharing characterized by monogamy inequality, one can
possibly quantify how much information an eavesdropper could
potentially obtain about the secret key to be extracted in quantum
cryptography~\cite{Paw}. In other words, the security of quantum
key distribution protocols that prohibits an eavesdropper from obtaining any
information without disturbance is guaranteed by MoE, the law of quantum
physics, rather than assumptions on the difficulty of
computation.

Here, we propose a new type of polygamy inequality for quantum
entanglement; in multi-party quantum systems, we first consider the
possible amount of bipartite entanglement distributed between a fixed party
and any subset of the rest parties. By using the summation of these
distributed entanglements, we provide an upper bound of the
distributed entanglement between a party and the rest. We then show that this upper bound
also plays as a lower bound of the general polygamy inequality of multi-party quantum entanglement;
therefore the strong polygamy of multi-party quantum
entanglement. For the case of multi-party pure states, we further show
that the strong polygamy of entanglement implies that
of quantum discord.

This paper is organized as follows. In Section~\ref{Sec: Bipartite
Quantum Correlations}, we briefly recall the definitions  and some
properties of bipartite quantum correlations such as entanglement of
assistance, quantum discord, one-way unlocalizable entanglement and
one-way unlocalizable  quantum discord. In Section~\ref{sub: spoly
entanglement}, we establish the strong polygamy of distributed
entanglement in terms of EoA, and we also show a close relation between the strong
polygamy of entanglement and quantum discord for multi-party pure states in Section~\ref{sub:
spoly discord}. In Section~\ref{Sec:Con}, we summarize our results.

\section{Bipartite Quantum Correlations}
\label{Sec: Bipartite Quantum Correlations}

For a bipartite quantum state $\rho_{AB}$, its one-way classical
correlation $\mathcal
{J}^\leftarrow(\rho_{AB})$ is
\begin{equation}\begin{split}
  \mathcal {J}^\leftarrow(\rho_{AB}) &=  \max_{\{M_x\}} \left[S(\rho_A)-\sum_x p_x S(\rho^x_A)\right],
\end{split}
\label{HVI}
\end{equation}
where $p_x\equiv \T[(I_A\otimes M_x)\rho_{AB}]$ is the probability
of the outcome $x$, $\rho^x_A\equiv \T_B[(I_A\otimes
{M_x})\rho_{AB}]/p_x$ is the state of system $A$ when the outcome
was $x$, and the maximum is taken over all the measurements
$\{M_x\}$ applied on system $B$~\cite{Henderson-Vedral01}.

For a tripartite pure state $\ket{\psi}_{ABC}$ with reduced density
matrices $\rho_{A}=\T_{BC}\ket{\psi}_{ABC}\bra{\psi}$, $\rho_{AB}
=\T_{C}\ket{\psi}_{ABC}\bra{\psi}$, and $\rho_{AC}
=\T_{B}\ket{\psi}_{ABC}\bra{\psi}$,
a trade-off relation between quantum entanglement and classical correlation was shown~\cite{KW}
\begin{align}
    S(\rho_A)&={\mathcal J}^\leftarrow(\rho_{AB})+E_f(\rho_{AC}),
    \label{KWmain1}
\end{align}
where
\begin{equation}
E_f(\rho_{AC})=\min \sum_{i}p_i S(\rho^{i}_{A}) \label{eof}
\end{equation}
is the {\em entanglement of formation}(EoF) of $\rho_{AC}$ ~\cite{BDSW},
whose minimization is taken over over all pure state decompositions
of $\rho_{AC}$,
\begin{equation}
\rho_{AC}=\sum_{i} p_i |\phi^i\rangle_{AC}\langle\phi^i|,
\label{decomp}
\end{equation}
with $\T_{C}|\phi^i\rangle_{AC}\langle\phi^i|=\rho^{i}_{A}$.

From the definition, $E_f(\rho_{AC})$ is considered as the minimum averaged entanglement needed
to prepare $\rho_{AC}$, and the term {\em formation} naturally arises. Furthermore, Eq.~(\ref{KWmain1})
can be interpreted as follows; for any tripartite pure state $\ket{\psi}_{ABC}$ (a three-party closed quantum system),
the total correlation between subsystems $A$ and $BC$ quantified by the entropy $S\left(\rho_A\right)$ consists of
the classical correlation ${\mathcal J}^\leftarrow(\rho_{AB})$ between subsystems $A$ and $B$, and the formation of entanglement $E_f(\rho_{AC})$
between $A$ and $C$.

As a dual quantity to EoF, the {\em entanglement of assistance} (EoA) is defined as the maximum average entanglement,
\begin{equation}
E_a(\rho_{AC})=\max \sum_{i}p_i S(\rho^{i}_{A}), \label{EoA}
\end{equation}
over all possible pure state decompositions of $\rho_{AC}$~\cite{LVvE03}.
EoA is clearly a mathematical dual to EoF because one takes the maximum average entanglement whereas the other
takes the minimum.

We also note that for a pure state $\ket{\psi}_{ABC}$, all possible pure state
decompositions of $\rho_{AC}$ can be realized by rank-1 measurements
of subsystem $B$, and conversely, any rank-1 measurement can be
induced from a pure state decomposition of $\rho_{AC}$~\cite{BGK}.
Thus $E_a(\rho_{AC})$ can be considered as the possible maximum average entanglement that can be distributed between $A$ and $C$ with
the assistance of the environment $B$. This makes the duality between EoF and EoA clearer because one is the formation of entanglement
whereas the other is the possible entanglement distribution.

Similarly to the duality between EoF and EoA, we have a dual quantity to $J^\leftarrow(\rho_{AB})$;
for a bipartite state $\rho_{AB}$,
the {\em one-way unlocalizable entanglement}~(UE) is defined as
\begin{equation}
\begin{split}
E_u^{\leftarrow}(\rho_{AB}) &:=  \min_{\{M_x\}} \left[S(\rho_A)-\sum_x p_x S(\rho^x_A)\right],\\
\end{split}
\label{UE}
\end{equation}
where the minimum is taken over all possible rank-1 measurements
$\{M_x\}$ applied on system $B$~\cite{BGK}.
Moreover, the trade-off relation in
Eq.~(\ref{KWmain1}) was also shown to have a dual relation in terms of EoA
and UE in three-party quantum systems~\cite{BGK}. For a three-party
pure state $\ket{\psi}_{ABC}$,
\begin{align}
S(\rho_A)&=E_u^{\leftarrow}(\rho_{AB})+E_a(\rho_{AC}).
\label{eq: 3UEEA}
\end{align}

For a bipartite state $\rho_{AB}$, quantum discord (QD) is defined as the difference between the mutual
information and one-way classical correlation~\cite{OZ,HV},
\begin{equation}
{\delta}^{\leftarrow}\left(\rho_{AB}\right)={\mathcal I}\left(\rho_{AB}\right)-{\mathcal J}^\leftarrow\left(\rho_{AB}\right),
\label{QD}
\end{equation}
where
\begin{equation}
{\mathcal I}\left(\rho_{AB}\right)=S\left(\rho_A\right)+S\left(\rho_B\right)-S\left(\rho_{AB}\right),
\label{mutual}
\end{equation}
is the mutual information of $\rho_{AB}$ with reduced density matrices $\rho_A$ and $\rho_B$ onto its subsystems $A$ and $B$
respectively.

Based on the duality between one-way classical correlation and UE, a dual quantity to QD was introduced; for a bipartite
quantum state $\rho_{AB}$ its one-way unlocalizable quantum discord (UD) is defined as~\cite{UD}
\begin{equation}
{\delta}_u^{\leftarrow}\left(\rho_{AB}\right)={\mathcal I}\left(\rho_{AB}\right)-E_u^\leftarrow\left(\rho_{AB}\right),
\label{UD}
\end{equation}
where $E_u^\leftarrow\left(\rho_{AB}\right)$ is the UE of $\rho_{AB}$ in Eq.~(\ref{UE}).

\section{Strong Polygamy of Quantum Correlations}
\label{Sec: Strong Polygamy}
\subsection{Strong Polygamy of Quantum Entanglement}
\label{sub: spoly entanglement}
In multi-party quantum systems, the distribution of bipartite entanglement quantified by EoA has
a polygamous relation as follows; for an $n+1$-party quantum state
$\rho_{AB_1\cdots B_n}$ with reduced density matrices
$\rho_{BA_i}$ on bipartite subsystems $AB_i$ for $i=1,\cdots,n$,
\begin{align}
E_a\left(\rho_{A(B_1\cdots B_n)}\right)
\leq& E_a\left(\rho_{AB_1}\right)+\cdots +E_a\left(\rho_{AB_n}\right)\nonumber\\
=&\sum_{i=1}^{n}E_a\left(\rho_{AB_i}\right),
\label{npolymixed}
\end{align}
where $E_a\left(\rho_{A(B_1\cdots B_n)}\right)$ is EoA of $\rho_{AB_1\cdots B_n}$
with respect to the bipartition between
$A$ and the rest, and $E_a\left(\rho_{BA_i}\right)$ is EoA of $\rho_{BA_i}$ for $i=1,\cdots,n$~\cite{gpoly}.

Let us denote $\B = \{B_1,\cdots,B_n \}$, that is, the set of
subsystems $B_i$'s, and consider a nonempty proper subset
$\X=\{B_{i_1},\cdots,B_{i_k}\}$ of $\B$ for $1 \leq k \leq n-1$.
Together with the complement $\X^c=\B-\X$ of $\X$ in $\B$,
$\rho_{AB_1\cdots B_n}$ can also be considered as a three-party
quantum state $\rho_{A\X\X^c}$. Furthermore, the polygamy inequality in~(\ref{npolymixed}) implies
\begin{align}
E_a\left(\rho_{A\B}\right)=&E_a\left(\rho_{A(\X\X^c)}\right)\nonumber\\
\leq& E_a\left(\rho_{A\X}\right)+E_a\left(\rho_{A\X^c}\right),
\label{polyAXXC}
\end{align}
where $E_a\left(\rho_{A\X}\right)$ and $E_a\left(\rho_{A\X^c}\right)$ are EoA of reduced density matrices
$\rho_{A\X}$ and $\rho_{A\X^c}$, respectively.
Because Inequality~(\ref{polyAXXC}) holds for any proper subset $\X$ of $\B$, we consider all possible nonempty proper subsets $\X$ of $\B$,
which lead us to the following inequality,
\begin{align}
E_a\left(\rho_{A\B}\right)&\leq \frac{1}{2^n-2}\sum_{\X}\left(E_a\left(\rho_{A\X}\right)+E_a\left(\rho_{A\X^c}\right)\right),
\label{3polyall}
\end{align}
where the summation is over all possible nonempty proper subsets $\X$'s.

Here we note that the set of all nonempty proper subsets of $\B$ is the same with the set of their complements;
\begin{align}
\{\X | \X \subset \B \}=\{\X^c| \X \subset \B \},
\label{sets}
\end{align}
thus we have
\begin{align}
\sum_{\X}E_a\left(\rho_{A\X^c}\right)=\sum_{\X}E_a\left(\rho_{A\X}\right),
\label{setsumequal}
\end{align}
and Eq.~(\ref{3polyall}) becomes
\begin{align}
E_a\left(\rho_{A\B}\right)\leq\frac{1}{2^{n-1}-1}\sum_{\X}E_a\left(\rho_{A\X}\right).
\label{spolyE}
\end{align}

For a nonempty proper subset $\X=\{B_{i_1},\cdots,B_{i_k}\}$ of $\B$ and its complement $\X^c=\{B_{i_{k+1}},\cdots,B_{i_n}\}$,
Inequality~(\ref{npolymixed}) also implies
\begin{align}
E_a\left(\rho_{A\X}\right)+&E_a\left(\rho_{A\X^c}\right)\nonumber\\
&\leq\sum_{j=1}^{k}E_a\left(\rho_{AB_{i_j}}\right)+\sum_{j=k+1}^{n}E_a\left(\rho_{AB_{i_j}}\right)\nonumber\\
&=\sum_{i=1}^{n}E_a\left(\rho_{AB_i}\right).
\label{upper}
\end{align}
By considering all possible nonempty proper subsets $\X$ of $\B$ and using Eqs.~(\ref{sets}) and (\ref{setsumequal}), we have
\begin{align}
\frac{1}{2^{n-1}-1}\sum_{\X}E_a\left(\rho_{A\X}\right)\leq \sum_{i=1}^{n}E_a\left(\rho_{AB_i}\right).
\label{upper2}
\end{align}

From inequalities~(\ref{spolyE}) and (\ref{upper2}), we have the following {\em strong polygamy inequalities}
of distributed entanglement in multi-party quantum systems;
for any multi-party state
$\rho_{AB_1\cdots B_n}$,
(pure or mixed)
\begin{align}
E_a\left(\rho_{A\B}\right)\leq&\frac{1}{2^{n-1}-1}\sum_{\X}E_a\left(\rho_{A\X}\right)\nonumber\\
\leq& \sum_{i=1}^{n}E_a\left(\rho_{AB_i}\right), \label{spolyE2}
\end{align}
where the first summation is over all nonempty proper subsets $\X$ of $\B = \{B_1,\cdots,B_n \}$.

Here, the term {\em strong} is twofold. First, Inequality~(\ref{spolyE2}) is in fact tighter than the usual polygamy inequality
in (\ref{npolymixed}). Moreover, we have considered the entanglement distribution (EoA) between the single party $A$ and all possible subsets $\X$'s
of $\B$ to obtain a tighter polygamy inequality whereas the usual polygamy inequality only considers EoA between $A$ and each single party ($B_i$'s)
in $\B$.

\subsection{Strong Polygamy of Quantum Discord}
\label{sub: spoly discord}

Let us now consider strong polygamy inequality of
quantum discord in multi-party quantum systems in terms of UD. We first note that the
definition of UD in Eq.~(\ref{UD}) and the relation between EU and
EoA in Eq.~(\ref{eq: 3UEEA}) lead us to the following relation
between ED and EoA; for a three-party pure state $\ket{\psi}_{ABC}$
with its reduced density matrices $\rho_{AB}$ and $\rho_{AC}$,
\begin{align}
E_a\left(\rho_{AB}\right)={\delta}_u^{\leftarrow}\left(\rho_{AC}\right)+S\left(\rho_{A|C}\right),
\label{EDEoA}
\end{align}
where $S\left(\rho_{A|C}\right)=S\left(\rho_{AC}\right)-S\left(\rho_{C}\right)$ is the conditional entropy of $\rho_{AC}$.
For a multi-party pure state
$\ket{\psi}_{A\B}=\ket{\psi}_{AB_1\cdots B_n}$ and a nonempty proper
subset $\X$ of $\B$, Eq.~(\ref{EDEoA}) implies
\begin{align}
E_a\left(\rho_{A\X}\right)={\delta}_u^{\leftarrow}\left(\rho_{A\X^c}\right)+S\left(\rho_{A|\X^c}\right),
\label{EDEoA2}
\end{align}
where $\rho_{A\X}$ and $\rho_{A\X^c}$ are the reduced density matrices of $\ket{\psi}_{A\B}$ on to subsystems $A\X$ and $A\X^c$, respectively.

Now we consider above equality for all possible nonempty proper subsets $\X$ of $\B = \{B_1,\cdots,B_n \}$ to obtain
\begin{align}
\sum_{\X}E_a\left(\rho_{A\X}\right)=&\sum_{\X}\left({\delta}_u^{\leftarrow}\left(\rho_{A\X^c}\right)+S\left(\rho_{A|\X^c}\right)\right)\nonumber\\
=&\sum_{\X}{\delta}_u^{\leftarrow}\left(\rho_{A\X^c}\right)+\sum_{\X}S\left(\rho_{A|\X^c}\right)\nonumber\\
=&\sum_{\X}{\delta}_u^{\leftarrow}\left(\rho_{A\X}\right)+\sum_{\X}S\left(\rho_{A|\X}\right),
\label{EDEoAsum}
\end{align}
where the last equality is due to Eq.~(\ref{sets}).
Furthermore, due to the complementary property of conditional entropy, we have
\begin{align}
S\left(\rho_{A|\X}\right)+S\left(\rho_{A|\X^c}\right)=0
\label{compcondent}
\end{align}
for any three-party pure state $\ket{\psi}_{A\X\X^c}$, and this implies
\begin{align}
\sum_{\X}S\left(\rho_{A|\X}\right)=0,
\label{compcondent2}
\end{align}
where the summation is over all nonempty proper subsets of $\B$.
From Eqs.~(\ref{EDEoAsum}) and (\ref{compcondent2}), we have
\begin{align}
\sum_{\X}E_a\left(\rho_{A\X}\right)=\sum_{\X}{\delta}_u^{\leftarrow}\left(\rho_{A\X}\right),
\label{sumEDEU}
\end{align}
for any multi-party pure state $\ket{\psi}_{A\B}$ and its reduced density matrix $\rho_{A\X}$.

Let us now consider UD of a bipartite pure state $\ket{\psi}_{A\B}$; the definition of UD in Eq.~(\ref{UD}) leads us to
\begin{align}
{\delta}_u^{\leftarrow}\left(\ket{\psi}_{A\B}\right)={\mathcal I}\left(\ket{\psi}_{A\B}\right)-E_u^\leftarrow\left(\ket{\psi}_{A\B}\right).
\label{UDpure}
\end{align}
For a bipartite pure state $\ket{\psi}_{A\B}$, we have
\begin{align}
{\mathcal I}\left(\ket{\psi}_{A\B}\right)=&S\left(\rho_A\right)+S\left(\rho_{\B}\right)-S\left(\ket{\psi}_{A\B}\right)\nonumber\\
=&2S\left(\rho_A\right),
\label{mutualpure}
\end{align}
thus Eq.~(\ref{UDpure}) becomes
\begin{align}
{\delta}_u^{\leftarrow}\left(\ket{\psi}_{A\B}\right)=2S\left(\rho_A\right)-E_u^\leftarrow\left(\ket{\psi}_{A\B}\right).
\label{UDpure1}
\end{align}

We note that any purification of $\ket{\psi}_{A\B}$ in
three-party quantum systems $A\B C$ is trivially a product state
$\ket{\psi}_{A\B}\otimes\ket{\phi}_C$ for some pure state
$\ket{\phi}_C$. From the definition of EU in Eq.~(\ref{UE}), we have
\begin{align}
E_u^\leftarrow\left(\ket{\psi}_{A\B}\right)=S\left(\rho_{A}\right)-E_a\left(\rho_{AC}\right),
\label{UE2}
\end{align}
where $\rho_{AC}$ is the reduced density matrix of
$\ket{\psi}_{A\B}\otimes\ket{\phi}_C$ on subsystems $AC$, which is
\begin{align}
\rho_A\otimes\ket{\phi}_C\bra{\phi}.
\label{redAC}
\end{align}
Because $E_a\left(\rho_{AC}\right)=0$ for the product state
$\rho_{AC}$, we have
\begin{align}
E_u^\leftarrow\left(\ket{\psi}_{A\B}\right)=S\left(\rho_{A}\right),
\label{UE3}
\end{align}
for the bipartite pure state $\ket{\psi}_{A\B}$, therefore
Eqs.~(\ref{UDpure1}) and Eq.~(\ref{UE3}) lead us
to
\begin{align}
{\delta}_u^{\leftarrow}\left(\ket{\psi}_{A\B}\right)=S\left(\rho_{A}\right).
\label{UDpure2}
\end{align}
We also note that EoA of $\ket{\psi}_{A\B}$ is just the entropy of
subsystems, thus
\begin{align}
{\delta}_u^{\leftarrow}\left(\ket{\psi}_{A\B}\right)=E_a\left(\ket{\psi}_{A\B}\right).
\label{UDEoA}
\end{align}

Now, from Eqs.~(\ref{sumEDEU}) and (\ref{UDEoA}) together with
Inequality~(\ref{spolyE}), we have
\begin{align}
{\delta}_u^{\leftarrow}\left(\ket{\psi}_{A\B}\right)\leq\frac{1}{2^{n-1}-1}\sum_{\X}{\delta}_u^{\leftarrow}\left(\rho_{A\X}\right),
\label{spolyD}
\end{align}
where the summation is over all non-empty proper subsets of $\B$.
In other words, strong polygamy inequality of entanglement in
(\ref{spolyE}) also implies the strong polygamy of quantum discord
for the case of multi-party pure states $\ket{\psi}_{A\B}$, that is,
closed quantum systems.

\section{Summary}
\label{Sec:Con}
We have proposed a strong polygamy inequality for multi-party quantum
entanglement; by considering the possible amount of entanglement distribution in terms of EoA
between a fixed party and any subset of the rest parties in a multi-party quantum system,
we have provided an upper bound of the distributed entanglement between a party and the rest. We have also shown that
this upper bound plays as a lower bound of the usual polygamy inequality.
We have further shown that the strong polygamy of entanglement implies that
of quantum discord for the case of multi-party pure states.

Our results strengthen the characterization of the polygamous nature of
entanglement in multi-party quantum systems. Moreover, our results shows a closed relation
between PoE and quantum discord, which provides a strong clue for possible relations
between PoE and other quantum correlation measures.
Noting the importance of the study on multipartite quantum correlations, our results can provide a rich reference for future
work on the study of quantum correlations in multi-party quantum systems.

\section*{Acknowledgments}
This research was supported by Basic Science Research Program through the National Research Foundation of Korea(NRF)
funded by the Ministry of Education, Science and Technology(2012R1A1A1012246).

\end{document}